# Toward a Dynamic Intellectual Property Protection Model in High-Growth SMEs

*Short Paper*

**Sam Pitruzzello**

School of Computing and Information Systems, University of Melbourne, Melbourne, Australia
s.pitruzzello@student.unimelb.edu.au

**Sean Maynard**

School of Computing and Information Systems, University of Melbourne, Melbourne, Australia
seanbm@unimelb.edu.au

**Atif Ahmad**

School of Computing and Information Systems, University of Melbourne, Melbourne, Australia
atif@unimelb.edu.au

## Abstract

*This paper addresses the challenges faced by High-Growth Small-to-Medium Enterprises (HG-SMEs) in balancing intellectual property (IP) protection with open innovation during periods of rapid growth. Despite developing valuable IP assets that drive success, HG-SMEs often struggle with cybersecurity concerns related to IP theft and data exfiltration amidst resource constraints and the competing demands of expansion. We examine the intersection of cybersecurity, IP protection and rapid scaling - an area currently underexplored in existing literature. Drawing on Dynamic Capabilities (DC), Knowledge-based View (KBV) and open innovation theoretical frameworks, we introduce a conceptual framework to guide HG-SMEs in effectively managing valuable IP assets. This research-in-progress paper outlines a qualitative methodology to validate and refine the model. By addressing the research question of how HG-SMEs manage cybersecurity to protect valuable IP assets, we aim to provide practical guidance for high-growth, technology-driven companies navigating the tension between robust IP protection and collaborative innovation.*

## Introduction

Australia has a rich history of creating innovative technologies as demonstrated by the success of the WAAAX companies (Wisetech, Altium, Afterpay, Appen and Xero). Collectively, these companies generate $3.4 billion in annual revenue, have a market capitalization of $116 billion and employ over 9,600 people. While this highlights the economic contributions of large technology firms, they started out as small businesses with high growth potential (Altium, 2024; Australian Stock Exchange, n.d.; Eyers, 2022; Bowen, 2024). To drive growth, high-growth small-to-medium enterprises (HG-SME) need intellectual property (IP) to develop innovative products, services and technologies (Krasniqi & Desai, 2016).

HG-SMEs' rapid expansion exposes them to a range of challenges and risks (Raby et al., 2022). Maintaining a focus on growth and commercial viability often overshadows cybersecurity concerns, despite substantial risks (Chidukwani et al., 2024; Olander et al., 2011). These problems are compounded by several factors unique to HG-SMEs - limited financial resources, lack of dedicated cybersecurity personnel (SMEs tend to have employees with multiple responsibilities), an over-reliance on IT outsourcing (Bada & Nurse, 2019; Chidukwani et al., 2024) and how to operate in an environment of open innovation while simultaneously protecting IP assets (Grimaldi et al., 2021; Ahmad et al., 2014). These factors make HG-SMEs vulnerable to cybersecurity risks from a variety of threat actors, including competitors, organized crime syndicates and nation-states who understand the value of IP (Kotsias et al., 2022). We define HG-SMEs as companies that experience high levels of growth, have the potential to scale globally and develop valuable IP assets.

Despite these challenges, there is limited research examining how HG-SMEs can effectively manage the tension between growth and protecting IP. Current research primarily focuses on cybersecurity implementation in large organizations (Bada & Nurse, 2019; Chidukwani et al., 2024) or discussing growth management in SMEs (Raby et al., 2022). Few studies examine the intersection of these domains. This paper addresses this gap by developing a conceptual framework that explains how HG-SMEs can effectively manage cybersecurity during rapid growth. We draw on two theoretical lenses, Dynamic Capabilities (DC) and Knowledge-based view of the firm (KBV) to explain the mechanisms through which organizations can balance these competing demands and propose a conceptual model that can be applied in practice.

For the preceding reasons, the research question is:

> *How do high-growth SMEs manage cybersecurity to protect their valuable IP?*

In the following section, the research methodology relating to the literature search is covered. The literature review and theoretical background on open innovation and knowledge management, IP protection in HG-SMEs, DC and KBV are then discussed. A conceptual framework is then presented which is built on the theoretical foundations of DC and KBV. The final section outlines the future research agenda and conclusion.

## Research Methodology

A literature review drawing from Webster & Watson (2002) and Okoli (2015) was conducted. The search and review provide an assessment of the latest research grouped under five core concepts and theoretical frameworks. The process of how articles were found, including the search terms, and analyzed is described in this section allowing the search to be repeatable. The first step involved a Google Scholar search, resulting in 5,509 articles. Google Scholar ranks search results by relevance therefore the first 250 articles were selected for review. These articles were first filtered by excluding theses, books and non-peer reviewed articles resulting in a short list of 68 articles. The list was extended to include high-quality articles in the reference list of short-listed papers, seminal works in entrepreneurship, innovation and cybersecurity and building on the author's past research. An additional 27 papers were included, providing a total of 95 papers. These papers were filtered by applying a second screen according to the following inclusion criteria - articles focusing on cybersecurity in SMEs, IP protection in SMEs, knowledge management from an IP protection perspective and the application of organizational theories to cybersecurity. This resulted in 46 papers that were analyzed for inclusion in the literature review. Table 1 on the following page summarizes the search process and results.

The literature review included in the theoretical background covers four areas. The first section discusses open innovation and knowledge management. An overview of IP protection in HG-SMEs follows. The third and fourth topics cover the theories and frameworks applied in this research – DC and KBV through the lens open innovation. Of the final 46 papers selected, 34 are included in the reference list with 21 contributing to the literature review and model development. The remaining 12 papers provide support in other areas of the paper. Following is a summary of the papers used in each section of the literature review. These articles form the foundation of the literature review and theoretical background that informed the design of the conceptual model. For details on the research papers used in the model, refer to Table 2 on page 6.

**Open Innovation and Knowledge Management (10 papers)**: Adriko & Nurse (2024), Ahmad et al. (2014), Barbero et al. (2011), Chesbrough (2003), Gassmann et al. (2010), Grimaldi et al. (2021), Hernández-Linares et al. (2021), Krasniqi & Desai (2016), Nonaka (1994), Raby et al. (2022)

**IP Protection in HG-SMEs (10 papers)**: Ahmad et al. (2019), Amara et al. (2008), Bada & Nurse (2019), Chesbrough (2003), Chidukwani et al. (2024), Grimaldi et al. (2021), Olander et al. (2011), Pitruzzello et al. (2017), Shedden et al. (2009), Shojaifar & Fricker (2023)

**Dynamic Capabilities & Knowledge-Based View (4 papers)**: Grant (1996), Hernández-Linares et al. (2021), Teece et al. (1997), Teece (2007)

| Table 1 – Literature search and filtering results | | |
|---|---|---|
| Search Date | Search String | Search Result |
| 28 Jan 25 | (cybersecurity OR cyber-attack OR "incident response" OR "cyber* incident") AND entrepreneur* ("small to medium business" OR "small to medium enterprise" OR SME) AND ("intellectual property" OR "intellectual property protection" OR "protect commercial* sensitive information" OR "protect commercial* sensitive data") | 2,330 |
| 28 Jan 25 | ("entrepreneurial" OR "entrepreneurship") AND ("SME" OR "small to medium business" OR "small to medium enterprise" OR "small to medium business" OR "SMB" OR "startup" OR "start-up") AND ("cybersecurity" OR "crises" OR "cyber-attack") AND ("incident response" OR "cyber incident" OR "cybersecurity incident" OR "cyber incident response") AND ("intellectual property protection" OR "IP protection" OR "protect commercial sensitive information" OR "protect commercial sensitive data" OR "commercial sensitive information theft" OR "commercial sensitive data theft") And ("Intellectual Property Protection" OR "IP Protection" OR "Knowledge Protection") AND ("small-to-medium enterprises" OR "SME" OR "SMEs") AND ("cybersecurity") | 949 1,270 |
| 3 Feb 25 | ("Intellectual Property Protection" OR "IP Protection") AND ("small-to-medium enterprises" OR "SME" OR "SMEs") AND (cybersecurity incident response) | 960 |
| | Total Papers found in Google Scholar Search | 5,509 |
| Filters | | |
| Filter Step 1 | Top search results from Google – most relevant/top ranked papers | 250 |
| Filter Step 2 | Remove books, theses and non-peer reviewed articles | 68 |
| | Additional papers selected from reference list of 68 short-listed papers | 27 |
| | Total papers short listed for detailed review | 95 |
| Filter Step 3 | **Short-Listed papers after filtering by final selection criteria** | **46** |

## Theoretical Background

In this section, four areas of literature are discussed. The first topic discusses open innovation and knowledge management. Secondly, the literature on protecting IP in HG-SMEs is covered. The third and fourth topics cover two theories selected – DC and KBV through the lens of open innovation. These theories are relied upon to develop a conceptual model.

### Open Innovation and Knowledge Management

Common themes relating to HG-SMEs centers on entrepreneurship (Krasniqi & Desai, 2016; Raby et al., 2022) and innovation (Barbero et al., 2011; Hernández-Linares et al., 2021). Entrepreneurship involves risk-taking and high levels of competitiveness with innovation intensity also a common characteristic of HG-SMEs (Hernández-Linares et al., 2021). Open innovation has been shown to deliver significant advantages over closed innovation and considered a key driver of growth (Gassmann et al., 2010; Grimaldi et al., 2021). First coined by Henry Chesbrough, open innovation is defined as "...a paradigm that assumes that firms can and should use external ideas as well as internal ideas, and internal and external paths to market..." (Chesbrough, 2003, p. 37). While closed innovation occurs entirely within an organization with little or no external input, open innovation requires a high degree of collaboration with external organizations.

Open innovation requires companies to form supply chains with a network of partnerships, suppliers, advisors and joint ventures (Chesbrough, 2003; Gassmann et al., 2010). Innovation and knowledge are closely linked and the inter-relationship between them has been a research focus for decades (Chesbrough, 2003; Gassmann et al., 2010; Nonaka, 1994). Open innovation can be considered a form of knowledge management since it requires the sharing of knowledge with external organizations in a considered and thoughtful manner (Ahmad et al., 2014; Nonaka, 1994). HG-SMEs rely heavily on IP assets to develop innovative products and gain a competitive advantage. While HG-SMEs understand the need to protect their valuable IP, engaging in open innovation creates tension since they risk losing their IP to competitors, ex-employees and other entities keen on stealing and exploiting valuable IP (Adriko & Nurse, 2024).

## IP Protection in HG-SMEs

IP can be defined as "the creation, ownership and control of original ideas as well as the representation of those ideas." (Whitman & Mattord, 2014, p. 52). IP needs to be protected since considerable time, effort and resources have been invested in its development. IP protection methods (IPPM) can be formal or informal. Formal methods are backed by law and include patents, trademarks, copyright and registered designs. Informal methods are not supported by laws and employ strategies such as design complexity, trade secrets and lead-time advantage (Amara et al., 2008; Olander et al., 2011).

IPPM can provide defense from IP infringement and cybersecurity practices can protect IP assets from sophisticated threat actors attempting to steal valuable IP (Ahmad et al., 2019; Pitruzzello et al., 2017; Shedden et al., 2009). However, IPPM can stifle open innovation if IP is controlled too tightly (Chesbrough, 2003; Grimaldi et al., 2021). This raises a challenge for HG-SMEs in terms of the type of IPPM and cybersecurity practices to implement and how to approach open innovation. While research suggests that SMEs have poor cybersecurity practices (Bada & Nurse, 2019; Chidukwani et al., 2024), technically sophisticated HG-SMEs understand the importance of implementing cybersecurity practices to protect not only their IP but the overall organization (Shojaifar & Fricker, 2023). Research also suggests that inappropriate implementation of IPPM can lead to suboptimal outcomes if the organization does not have the right expertise (Grimaldi et al., 2021).

In summary, HG-SMEs need to implement sound cybersecurity practices to protect their IP assets from malicious actors while also engaging in open innovation to drive innovation and growth. If implemented diligently, open innovation can strike the right balance between IP protection and maximizing the value of innovative products and services. HG-SMEs can increase their chances of success by approaching these challenges with a flexible and dynamic mindset.

## Dynamic Capabilities (DC)

DC states that firms need to adapt their resources with changing market and economic conditions. The ability to reconfigure and transform assets to gain or maintain a competitive advantage is considered a dynamic capability (Teece et al., 1997). DC builds on the Resource-based View (RBV) of the firm, considered a 'static' view or theory (Hernández-Linares et al., 2021) and allows firms the "ability to integrate, build and reconfigure internal and external competences to address rapidly changing environments" (Teece et al., 1997, p. 516). Teece (2007) enhanced his original DC theory by introducing two micro-foundations to support resource and organizational transformation – sensing opportunities and threats and seizing opportunities (Teece, 2007). The complex nature of cybersecurity and protecting IP during rapid organizational growth requires a novel approach that addresses both the dynamic nature of these challenges with the competing demands that HG-SMEs face. Therefore, DC is a suitable theoretical framework to apply to HG-SMEs. RBV has also been applied and extended to another important theoretical framework – KBV (Grant, 1996) discussed in the next section through the lens of open innovation.

## Knowledge-Based View Through the Lens of Open Innovation

KBV states that knowledge, both explicit and tacit, is held by individuals and leveraged by firms for productive output and competitive advantage (Grant, 1996). KBV is related to DC except it focuses specifically on knowledge and in the case of tacit knowledge, that which is stored in a person's mind as 'know how'. This raises an important question - who owns tacit knowledge and how do organizations retrieve, store and transmit tacit knowledge to maximize its value? Grant (1996) suggests that companies need to implement clauses in employment agreements that assign IP developed by a person while working at the firm to the company (Grant, 1996). The storage and retrieval of knowledge has been covered in knowledge management research with its application in information systems a core concept (Nonaka, 1994). Open innovation is the process of sharing knowledge and IP with external parties to accelerate the commercialization of innovations and maximize the value of IP. HG-SMEs need to implement sound cybersecurity practices to protect their IP assets from malicious actors while also engaging in open innovation to drive innovation and growth. If implemented diligently, open innovation can strike the right balance between IP protection and maximizing the value of innovative products and services.

# Conceptual Model

Figure 1 below illustrates the conceptual model - Dynamic IP Protection (DIPP). DIPP integrates DC, KBV and OI. The model arranges activities in a matrix with the theoretical frameworks forming the columns and the two operating modes – closed IP development and open innovation in two rows. The four components of DC and KBV are – i) sense opportunities and threats and knowledge absorption, ii) seize opportunities and absorb knowledge, iii) knowledge transfer and iv) transformation and appropriate knowledge. Activities run across either row depending on whether the HG-SME is operating in closed IP development mode or open innovation mode. In closed development mode, the HG-SME is working towards commercializing IP and developing innovative products and services. In open innovation mode, the HG-SME is collaborating with external organizations to explore new ideas. The arrows indicate the flow and order of each activity. Both rows lead the updating IPPM and the ultimate outcome – developing the capability to build a secure, resilient firm that protects its valuable IP from cybersecurity threats.

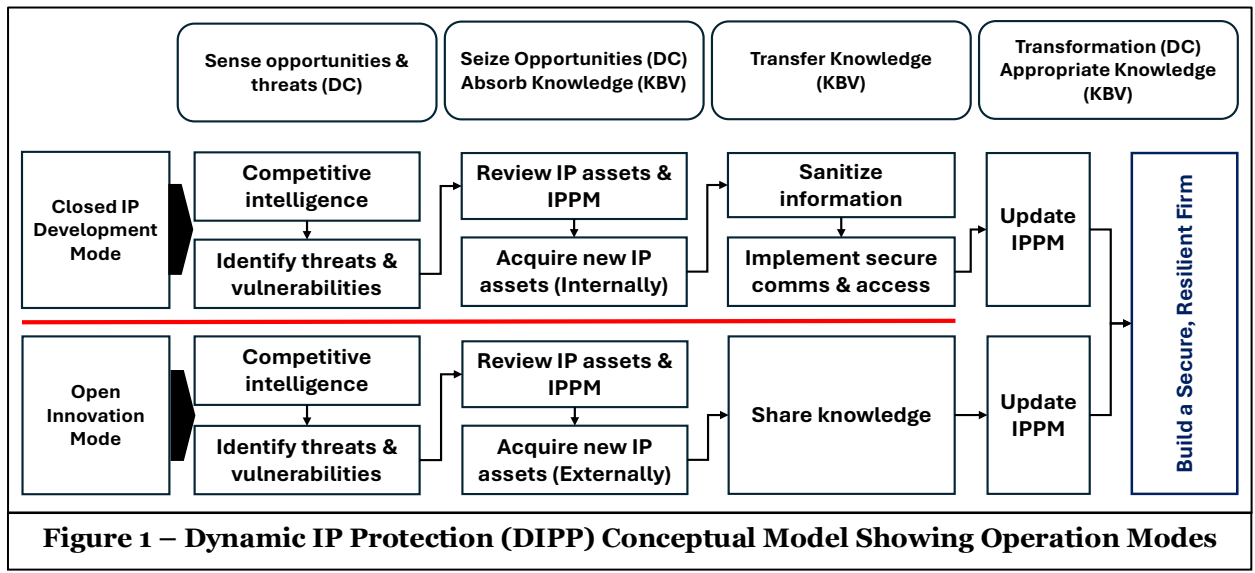

**Figure 1 – Dynamic IP Protection (DIPP) Conceptual Model Showing Operation Modes**

HG-SMEs operate in challenging, highly uncertain and constantly changing environments leading to high failure rates compared to large mature companies. In addition, HG-SMEs have limited resources in terms of time, resources and funds. As a result, the DIPP model has been chosen as a suitable framework as it is anticipated to be easy to implement, intuitive and align with most SME business operations. The theoretical lenses chosen to develop the model, DC and KBV, have been selected for several reasons. First, they cover knowledge concepts that have been applied in organizations. Secondly, they are easy to apply in practice. Finally, they have been widely applied in organizational research particularly in the fields of entrepreneurship, innovation and cybersecurity, all of which are relevant to this research project (Ahmad et al., 2014; Grant, 1996; Teece et al., 1997). The following sections describe each part of the DIPP model and the activities within each section. Table 2 below summarizes the constructs used in the conceptual research model. Each construct is defined along with the relevant references.

## *Sense Opportunities and Threats (DC)*

Sensing opportunities and threats, a core micro-foundation of Teece's DC, involves the identification of new opportunities and threats (Teece, 2007). In this research, threats relate to cybersecurity threats and vulnerabilities, in particular threats to HG-SMEs' IP assets. Activities are the same in both close IP development mode and open innovation mode – collect competitive intelligence and identify threats and vulnerabilities. To compete effectively in their markets, HG-SMEs need to obtain competitive intelligence on competitors while denying them access to their IP. Following the competitive intelligence activity, HG-SMEs need to assess threats and vulnerabilities. This requires gathering large amounts of information, analyzing the information and making sense of it (Ahmad et al., 2019).

| Table 2 – Definitions of constructs in the conceptual model | | |
|---|---|---|
| **Construct** | **Definition** | **References** |
| Sense | In the first part of the model, HG-SMEs need to sense opportunities and threats by collecting competitive intelligence and identifying cybersecurity threats and vulnerabilities to their IP assets. This involves gathering and analysing large amounts of information. | Ahmad et al. (2019), Teece et al. (2007), NIST CSF (2024) |
| Seize & Absorb | Seizing opportunities and absorbing knowledge requires HG-SMEs review IPPM and acquiring or developing new IP assets either internally or with external partners. | Amara et al., (2008), Gassmann et al. (2010), Grant (1996), Teece et al. (2007) |
| Transfer Knowledge | HG-SMEs switch between closed IPP mode and open innovation mode. When closed, knowledge needs to be secured by sanitizing information and securing communications. In open innovation mode, the organization switches to open knowledge management and information systems. | Ahmad et al. (2014), Grant (1996) |
| Transform & Appropriate | HG-SMEs are now ready to transform the organization by appropriating knowledge to derive value and reconfigure capabilities to maintain competitive advantage. This requires commercializing IP in a secure manner with key IPPM activities. | Ahmad et al. (2019), Grant (1996), Grimaldi et al. (2021), Teece (2007) |

Industry standards and frameworks can be useful in practice and applicable in this part of the model. For example, assessing threats aligns to risk assessment recommendations covered in ISO 27001 in section 'Information Security Risk Assessment' where it states, "The organization shall perform information security risk assessments at planned intervals or when significant changes are proposed..." (International Organization for Standardization, 2013, p. 7). Similarly, the NIST cybersecurity framework outlines several recommendations for assessing risks by identifying vulnerabilities in assets, receiving cyber threat intelligence from external sources and understanding the impacts and likelihood of threats exploiting vulnerabilities (National Institute of Standards and Technology, 2024). HG-SMEs need to ascertain whether their portfolio of IP assets is adequate in light of their competitive intelligence discoveries, which leads onto the next part of the model – seizing opportunities and absorbing knowledge.

### Seize Opportunities (DC) and Knowledge Absorption (KBV)

Seizing opportunities and knowledge absorption are similar concepts from DC and KBV respectively and have been grouped together in the model (Teece, 2007). DC suggests that firms need to seize new opportunities once they have been identified. This is similar to knowledge absorption in KBV that requires organizations to acquire and integrate new knowledge into their operations (Grant, 1996). The activities are the same in both closed IP development mode and open innovation mode with one difference – whether acquisition or development is carried out internally or externally.

The first activity involves reviewing IP assets and IPPM. ISO 27001 can inform IPPM reviews recommending "The organization shall evaluate the information security performance and the effectiveness of the information security management system" (International Organization for Standardization, 2013, p. 7). There are several IPPM that HG-SMEs can implement including patents, registered designs, trademarks and copyright protection. Implementing appropriate IPPM depends on the HG-SMEs operating environment, regulatory requirements and the nature of their innovations (Amara et al., 2008). For example, biotechnology firms are highly regulated and invest heavily in research and development therefore patents are used extensively to protect new medical innovations. Conversely, software firms operating in low regulatory environments have lower compliance costs and since software generally cannot be protected with patents, copyright and registered designs are the most common type of IPPM. The second activity in the model, acquisition or development of new IP assets, requires significant time, resources and funding. This can be a limiting factor for HG-SMEs with open innovation providing an expedient way to acquire new assets. When partnering with external organizations, it is important that HG-SMEs implement sound, fair and clear contractual agreements to protect all parties (Gassmann et al., 2010).

### Knowledge Transfer (KBV)

Knowledge transfer activities allow the HG-SME to switch from closed IP protection mode to open innovation mode. To build competitive advantage, HG-SMEs need to transfer knowledge throughout the organization efficiently (Grant, 1996). Diligence is required when sharing or allowing access to sensitive IP during its development. Therefore, we propose the closed IP development mode where knowledge is securely transferred within the organization and tightly controlled from external parties. The two activities in this mode are sanitizing knowledge and implementing secure communications and access. Sanitizing knowledge involves converting tacit knowledge to codified knowledge for storage, anonymizing personally identifiable information and encrypting sensitive information. Secure communications and access to information systems involve encrypting information for secure storage and communication and implementing access control policies including role/rule-based access and zero-trust policies. In open innovation mode, the organization switches to the traditional knowledge management function of sharing knowledge using clear, open and easy to access to information systems (Ahmad et al., 2014). These activities along with the review of IPPMs, lead to the ultimate outcome for HG-SMEs discussed in the next section.

### Transformation (DC) and Knowledge Appropriation (KBV)

In Grant's (1996) KBV, knowledge appropriation refers to an organization deriving value from knowledge assets (Grant, 1996). It is analogous to transformation in Teece (2007) DC framework where organizations reconfigure their assets and capabilities to transform their organizations to maintain a competitive advantage (Teece, 2007). This requires HG-SMEs to commercialize their IP assets and release innovative products and services in a secure manner by implementing sound IPPM. There is one key activity in this part of the model – updating IPPMs. This activity is the same in both closed IP development mode and open innovation mode. Finally, the model illustrates the ultimate outcome and its key purpose of developing capability within HG-SMEs – building a secure, resilient firm. In addition to the IPPMs discussed above, activities such as monitoring the market for IP infringements and conducting cybersecurity tests eliminate information exfiltration should also be conducted in this step (Ahmad et al., 2019; Grimaldi et al., 2021).

## Conclusion and Future Research

During growth phases, HG-SMEs experience rapid changes. This places them in challenging situations in uncertain environments and often operating with limited resources. In addition, HG-SMEs find themselves balancing the need to innovate in an open environment while protecting valuable IP. This research-in-progress paper proposes a model that explains how HG-SMEs can transform their cybersecurity practices and IPPM. Drawing from DC, KBV and open innovation, four key steps were applied in developing the model – sensing opportunities and threats, seizing opportunities and absorbing knowledge, knowledge transfer and knowledge appropriation and transformation. The aim of this research is to validate the model to ensure its suitability for HG-SMEs that need to protect valuable IP. This research will address a critical gap, since there has been insufficient research that addresses the cybersecurity needs of smaller entrepreneurial organizations. From a theoretical perspective, DC, KBV and open innovation have been integrated for a key reason – it will inform a 'mode change' where HG-SMEs lock down communications and access to sensitive information during IP development.

### Future Research

This research will be based on qualitative research methods with multiple case studies. Qualitative research has been chosen for this research project for several reasons. Yin (2013) compares the various methods available to researchers based on three criteria – a) the form of research question, b) whether the researcher requires control over events and c) does the research focus on contemporary events (Yin, 2017, p. 9). Surveys are normally quantitative in nature and set out to answer 'who, what and where' type questions. The research question in this project is a 'How?' type question thereby limiting the choices to experiment, historical and case study. This research is not studying history, instead it is focused on contemporary issues so the historical study option can be eliminated. This leaves experiment and case study. Experiments require the researcher to have some control over variables, tests or subjects. This is not the case in this research study therefore a case study approach is best suited (Yin, 2017).

Eight to twelve case studies are recommended as an adequate number to answer a research question and develop a model (Yin, 2017). Up to four representatives from each SME will be interviewed using a semi-structured interview process. The individuals selected for participation in the case study will be senior leadership executives, founders and board directors. Collectively, the individuals in each case study will have thorough insight and knowledge across the entire organization since they are involved in building and operating the company. Following the development of the model, focus group workshops will be run to validate the model. Each case study participant will be invited to participate in the workshops. On completion of the workshops, the feedback gathered will be used to validate and refine the model. Case study participants will be selected according to the following criteria:

1. The case unit is the organization (HG-SMEs)
2. Case units share similar structure (size = SME) and located in Australia (location: AU)
3. Case study participants develop valuable IP and innovative technologies
4. Case study participants deliver products and services to customers in a range of industries

This research will contribute to the body of knowledge on cybersecurity and IP protection in SMEs with a particular focus on high-growth entrepreneurial SMEs.